\documentclass[10pt,journal,twocolumn]{IEEEtran}
\usepackage{filecontents}
\usepackage[noadjust]{cite}
\usepackage[labelsep=period]{caption}
\usepackage{amsfonts}
\usepackage[dvips]{color, graphicx}
\usepackage[cmex10]{amsmath}
\usepackage{pgfplotstable}
\usepackage{array}
\usepackage{booktabs}
\usepackage{times}
\usepackage[normalem]{ulem} 

\ifCLASSOPTIONcompsoc
\usepackage[caption=false,font=normalsize,labelfont=sf,textfont=sf]{subfig}
\else
\usepackage[caption=false,font=footnotesize]{subfig}
\fi

\DeclareMathOperator*{\E}{E}

\long\def\symbolfootnote[#1]#2{\begingroup%
\def\thefootnote{\fnsymbol{footnote}}\footnote[#1]{#2}\endgroup}

\begin{document}
\title{Channel Estimation with \\ Systematic Polar Codes}
\author{
Liping~Li,~\IEEEmembership{~Member,~IEEE}, Zuzheng~Xu,~\IEEEmembership{~Student Member,~IEEE,} Yanjun~Hu~\IEEEmembership{~Senior Member,~IEEE}
\thanks{This work was supported in part by National Natural Science Foundation
of China through grant 61501002, in part by Natural Science Project of
Ministry of Education of Anhui through grant KJ2015A102, in part by Talents
Recruitment Program of Anhui University.

The authors are with the Key Laboratory of Intelligent Computing and Signal Processing
of the Ministry of Education of China,
Anhui University, China (Email: liping\_li@ahu.edu.cn)}
}

\maketitle
\begin{abstract}
Study of polar codes in fading channels is of great importance when applying polar codes in wireless communications.
Channel estimation is a fundamental step for communication to be possible in fading channels.
For both systematic and non-systematic polar codes, construction of them is based on an information set and
the known frozen bits. Efficient implementation
of systematic and non-systematic polar codes exists.
When it comes to channel estimation or channel tracking, additional
pilot symbols are inserted in the codeword traditionally.
In this paper, to improve the performance of polar codes in the finite domain,
pilot symbols are selected from the coded symbols themselves.
In order to keep the existing efficient structure of polar code
encoding, pilot selection is critical since not all selections can reuse the existing structure.
In this paper, two pilot selections denoted as Uneven Pilot Selection (UEPS) and
Even Pilot Selection (EPS) are proposed, which do not change the efficient polar encoding structure.
The proposed UEPS and EPS is proven to satisfy the efficient construction condition.
The performance of EPS is shown in this paper to outperform both the UEPS and the traditional pilot insertion scheme.
Simulation results are provided which verify the performance of the proposed pilot selection schemes.
\end{abstract}

\section{Introduction}\label{sec_ref}
Polar codes \cite{arikan_iti09}, due to the low encoding and decoding complexity of $\mathcal{O}(N \log N)$
and the capability to realize the capacity of binary-input discrete memoryless channels, are of great
potential for future communication applications. The construction of polar code is
to select {the} $K$ best bit channels among $N$ bit channels to convey information. Besides
the {binary erasure channel (BEC)}, polar code construction in all other channel types do not have
iterative equations and therefore has a high computation complexity
\cite{arikan_iti09,mori_icl09,vardy_it13,trifonov_itc12,wu_icl14}.

To improve the error performance of polar codes in the finite domain,
{successive cancellation list (SCL)} decoding is
proposed in \cite{niu_itc13}\cite{vardy_it15} and belief {propagation} (BP) decoding is applied to polar codes in
\cite{arikan_icl08}\cite{urbanke_isit09}\cite{guo_isit14}.
Another direction in improving the error performance of polar codes is to encode
systematically instead of the original non-systematic encoding \cite{arikan_icl11}.
Systematic polar codes are shown
to outperform the non-systematic polar codes in terms of the BER performance with almost no extra cost in the decoding
process. Therefore, systematic polar codes are the focus of this paper.

When employing polar codes in a wireless communication scenario, channel state information (CSI) needs to
be estimated before further communication is possible. In OFDM and MIMO systems, pilot-aided channel estimation
and tracking is studied \cite{li_tvt00,mengali_tsp01,li_twc02,gibson_twc05,li_tc06,gershman_tsp06}.
Least square (LS) and minimum mean square error (MMSE) channel estimators are commonly used in these works,
which are also the estimators in this paper. Pilots are inserted either in the frequency domain
or the time domain or in both domains as the LTE systems \cite{3gpp_13}.

In this paper, pilots are not inserted as traditionally done. Instead, pilots are selected from
the coded symbols.
The motivation behind this pilot selection is to improve the performance of
polar codes in the finite domain. The selected pilots not only serve the purpose of channel estimation
or tracking,
but also provide stronger protection to the information bits during the decoding processing.
However, as systematic polar codes can be
constructed or implemented efficiently \cite{li_socc15}\cite{gross_tc16}, the pilot selection scheme
in principle should not alter the existing efficient encoding structure.
Note that binary phase shift keying (BPSK) is the modulation scheme in this work. Therefore, we
interchangeably use pilot symbols, pilot bits, or coded symbols without further noticing.

Let $G_N$ be the generator matrix of the polar code with the block length $N$, and $\mathcal{A}$
be the set containing the indices of the information bit channels. The submatrix
$G_{\mathcal{AA}}$ is taken from the matrix $G_N$ with rows and columns both specified by the
the set $\mathcal{A}$.
In this paper, the
efficient encoding condition is presented both in our matrix form \cite{li_socc15}:
$G_{\mathcal{AA}}=G_{\mathcal{AA}}^{-1}$
and in the domination contiguous form in
\cite{gross_tc16}: the set $\mathcal{A}$ being domination contiguous ({defined in (\ref{eq_domination})).
We prove that the matrix form $G_{\mathcal{AA}}=G_{\mathcal{AA}}^{-1}$ is equivalent to the set $\mathcal{A}$
being domination contiguous.

Based on the efficient encoding condition, two pilot selection schemes are proposed: the uneven pilot selection (UEPS)
and the even pilot selection (EPS).
With pilots selected from the coded symbols, the new encoding set is
$\mathcal{C} = \mathcal{A} ~\cup ~\mathcal{P}_f$, where
$\mathcal{P}_f$ is the set containing pilot
symbols selected from the frozen set $\mathcal{\bar{A}}$.
The two proposed selections UEPS and EPS are proven
to still meet the efficient encoding condition of $G_{\mathcal{CC}}=G_{\mathcal{CC}}^{-1}$ and the set
$\mathcal{C}$ is proven to be still domination contiguous.
The efficiency and decoding performance of the proposed pilot selections are analyzed in the paper.
The decoding performance of EPS is analyzed and shown by simulations to be better than the traditional
pilot insertion scheme.

The contributions of the paper can be summarized as:
1) The existing two efficient encoding conditions for systematic polar codes \cite{li_socc15}\cite{gross_tc16} are proven to be equivalent;
2) Pilots are selected from the coded symbols instead of being inserted additionally. Two pilot selection schemes are proposed which meet the efficient encoding conditions;
3) Theoretical and numerical results are provided which show that the proposed pilot selection scheme outperforms the traditional pilot insertion scheme.

The main notations in this paper are {firstly} introduced below.
A row vector with elements $(v_1,v_2,...,v_N)$ is written as $v_1^N$.
Given a vector $v_1^N$, the
vector $v_i^j$ is a subvector $(v_i, ..., v_j)$ with $1 \le i,j \le N$. If there is a set $\mathcal{A}
~{{\subseteq}} ~\{1,2,...,N\}$,
then $v_{\mathcal{A}}$ denotes a subvector with elements in $\{v_i, i \in \mathcal{A}\}$.

The rest of the paper is organized as follows. Section \ref{sec_background} is on the basics of polar codes.
It is proven in Section {\ref{sec_efficient_construction}}
that the two efficient
encoding conditions are essentially the same. In Section \ref{sec_pilot_selection}, two pilot selection schemes
are presented and proven to be efficiently encodable. The efficiency and the decoding performance is also analyzed
in this section. Simulation results are provided in Section \ref{sec_simulation}. Concluding remarks are presented in the
last section.


\section{Polar Code Background}\label{sec_background}
The polarization of $N$ independent underlying channels $W$ is realized through two stages: channel combining and splitting.
Let $W(y|x)$ be the transition probability of $W$. Let $G_N$ be the generator matrix of the polar code
with a block length $N$. In \cite{arikan_iti09}, the generator matrix is $G_N=BF^{\otimes n}$ where $F=\left[\begin{smallmatrix} 1&0 \\ 1&1 \end{smallmatrix}\right]$,  $B$ is a bit-reversal permutation matrix, and
$F^{\otimes n}$ means the $n$th Kronecker power of the matrix $F$ in the binary field.
Denote $\mathcal{X}$ as the alphabet set of the input $x$.
The channel combining stage produces a vector channel {$W_N$ defined as}
\begin{equation} \label{eq_w_vector}
W_N(y_1^N|u_1^N) = W^N(y_1^N | u_1^N G_N)
\end{equation}
where $W^N(y_1^N |u_1^N G_N) = W^N(y_1^N|x_1^N) = \prod_{i=1}^N W(y_i|x_i)$.

This vector channel can then be split into $N$ bit channels \cite{arikan_iti09}:
\begin{equation}\label{eq_wni}
W_N^{(i)}(y_1^N,u_1^{i-1} | u_i) = \sum_{u_{i+1}^N \in \mathcal{X}^{N-i}}\frac{1}{2^{N-1}}W_N(y_1^N|u_1^N)
\end{equation}
Note that the summand in (\ref{eq_wni}) is conditioned on $u_1^N = (u_1^i, u_{i+1}^N)$ and is summed
over $u_{i+1}$ to $u_N$.
The channel with the transition probability $W_N^{(i)}(y_1^N,u_1^{i-1}|u_i)$ is the channel
that the source bit $u_i$ goes through; it is
called bit channel $i$. In the following of the paper, we use $W_N^{(i)}$ to refer to bit channel $i$.
Channels are polarized after these two stages in the sense that bits transmitting in these bit channels either experience almost noiseless channels or almost completely noisy channels when  $N$ is large. The idea of polar codes is to transmit information bits on those noiseless channels. The fixed bits are made known to both the transmitter and receiver.

Mathematically, the encoding is a process to obtain the encoded bits ${x}_1^N$
through ${x}_1^N = {u}_1^N G_N $ for a
given source vector ${u}_1^N$. The source vector ${u}_1^N$ consists of the information bits and the frozen bits, denoted by ${u}_{\mathcal{A}}$ and ${u}_{\bar{\mathcal{A}}}$, respectively. Here the set $\mathcal{A}$ includes the indices for the information bits and $\bar{\mathcal{A}}$ is the complimentary set.
Both sets $\mathcal{A}$ and $\mathcal{\bar{A}}$ are in
$\{1,2,...,N\}$ for polar codes with a block length $N=2^n$. The bit channels in $\mathcal{A}$ are
better than those in $\mathcal{\bar{A}}$. Or in other words, the bit channels in $\mathcal{\bar{A}}$
should be stochastically degraded with respect to those in $\mathcal{A}$.

{For two bit channels $i$ and $j$, denote $W_N^{(j)} \preceq W_N^{(i)}$ if bit channel $j$
is stochastically degraded with respect to bit channel $i$ as \cite{vardy_it13}. In mathematical terms, the
information  set $\mathcal{A}$ for the underlying channel $W$ has the following property:
\begin{equation}\label{eq_A}
\mathcal{A} = \{i \in \{1,2,...,N\} | W_N^{(j)} \preceq W_N^{(i)}, ~ j \in \mathcal{\bar{A}} \}
\end{equation}

Denote the size of $\mathcal{A}$ as $K=|\mathcal{A}|$.
In this paper, suppose the set $\mathcal{A}$ is found from calling any sorting algorithm such as \cite{vardy_it13}.
The frozen bits in $u_{\mathcal{\bar{A}}}$ are fixed bits which are made known to the receiver.

The systematic polar code \cite{arikan_icl11} is constructed by
specifying a set of indices of the codeword $\mathbf{x}$ as the indices to convey the information bits.
Denote this set as $\mathcal{B}$ and the complementary set as $\bar{\mathcal{B}}$.
The codeword $\mathbf{x}$ is thus split as
$({x}_{\mathcal{B}},{x}_{\bar{\mathcal{B}}})$. With some manipulations, we have
\begin{equation} \label{eq_xb_xbc}
({x}_{\mathcal{B}},{x}_{\bar{\mathcal{B}}}) = ({u}_{\mathcal{A}}G_{\mathcal{AB}}+{u}_{\bar{\mathcal{A}}}G_{\bar{\mathcal{A}}\mathcal{B}}, {u}_{\mathcal{A}}G_{\mathcal{A\bar{B}}}+{u}_{\bar{\mathcal{A}}}G_{\mathcal{\bar{A}\bar{B}}})
\end{equation}
The matrix $G_{\mathcal{AB}}$ is a submatrix of the generator matrix with elements
$\{G_{i,j}\}_{i \in \mathcal{A}, j \in \mathcal{B}}$. The vector $u_{\mathcal{A}}$ can be obtained as the following
\begin{equation} \label{eq_uacd}
u_{\mathcal{A}} = (x_{\mathcal{B}}-u_{\bar{\mathcal{A}}}G_{\mathcal{\bar{A}B}})(G_{\mathcal{AB}})^{-1}
\end{equation}
From (\ref{eq_uacd}), it is seen that $x_{\mathcal{B}} \mapsto u_{\mathcal{A}}$ is one-to-one if the following
two conditions are met:
\begin{eqnarray} \label{eq_sys_con_1}
&&x_{\mathcal{B}} {~\text{has the same elements as~}} u_{\mathcal{A}} \\ \label{eq_sys_con_2}
&&G_{\mathcal{AB}} {~\text{is invertible}}
\end{eqnarray}
In \cite{arikan_icl11}, it is shown that $\mathcal{B} = \mathcal{A}$ satisfies  these two conditions in order to establish the
one-to-one mapping $x_{\mathcal{B}} \mapsto u_{\mathcal{A}}$. In the rest of the paper, the systematic encoding of polar
codes adopts this selection of $\mathcal{B}$ to be $\mathcal{B} = \mathcal{A}$. Therefore we can rewrite (\ref{eq_xb_xbc}) as
\begin{equation} \label{eq_xb_xbc_2}
({x}_{\mathcal{A}},{x}_{\bar{\mathcal{A}}}) = ({u}_{\mathcal{A}}G_{\mathcal{AA}}+{u}_{\bar{\mathcal{A}}}G_{\bar{\mathcal{A}}\mathcal{A}}, {u}_{\mathcal{A}}G_{\mathcal{A\bar{A}}}+{u}_{\bar{\mathcal{A}}}G_{\mathcal{\bar{A}\bar{A}}})
\end{equation}

{\emph{\textbf{Remark:~}}}In the context of the systematic polar codes, it is convenient to
refer to the generator matrix $G_N$ as
the one without permutation, namely $G_N=F^{\otimes n}$. The equation (\ref{eq_xb_xbc_2}) is established
under this matrix $G_N$ without permutation. From now on, the matrix $G_N$ is in this form without the permutation
matrix $B$ unless stated otherwise.

The successive cancellation (SC) decoding of polar codes is proposed in \cite{arikan_iti09},
which has a low complexity of
$\mathcal{O}(N \log N)$. The decision statistic of the SC decoder is:
\begin{eqnarray}
\hat{u}_i = \bigg\{
\begin{array} {cr}
u_i, & \text{if~~} i \in \mathcal{\bar{A}} \\
h_i(y_1^N, \hat{u}_1^{i-1}), & \text{if~~} i \in \mathcal{A}
\end{array}
\end{eqnarray}
where
\begin{eqnarray}
h_i(y_1^N, \hat{u}_1^{i-1}) = \bigg\{
\begin{array}{cl}
0, &\text{if~~} \frac
{W_N^{(i)}(y_1^N, \hat{u}_1^{i-1}|0)}{W_N^{(i)}(y_1^N, \hat{u}_1^{i-1}|1)} \ge 1 \\
1, & \text{otherwise~~}
\end{array}
\end{eqnarray}
The bits are decoded in the order from 1 to $N$. One bit error in $\hat{u}_i$ will
propagate to the information bit $j$ with $j > i$.
This results in an non-satisfactory
performance of polar codes in the finite domain
\cite{urbanke_isit09,eslami_allerton10, eslami_isit11,guo_isit14}.

\section{Efficient Construction of Polar Codes}\label{sec_efficient_construction}
The systematic mapping in (\ref{eq_xb_xbc_2}) can be simplified as in \cite{li_socc15} from the fact
that $G_{\mathcal{\bar{A}A}} = \mathbf{0}$.
Here we provide another proof of $G_{\mathcal{\bar{A}A}} = \mathbf{0}$
which is a simplified version of the one cited from \cite{li_socc15}.
\newtheorem{proposition}{Proposition}
\begin{proposition}\label{proposition_Gaca}
For a polar code with an information set $\mathcal{A}$ as defined in (\ref{eq_A}),
the submatrix $G_{\mathcal{\bar{A}A}}$ of $G_N$ is a zero matrix: $G_{\mathcal{\bar{A}A}}= \mathbf{0}$.
\end{proposition}
\begin{IEEEproof}
Let $\langle i-1\rangle_2 = (i_1,i_2,...,i_n)$ be the binary expansion of {the index of}
bit channel $i$ ($ 1 \le i \le N$).
The bit channel $i$ and $j$ is said to have a binary domination relation if
\begin{equation}\label{eq_domination}
i \succeq j {~~\text{iff for all~~} 1 \le t \le n,~~ i_t \ge j_t}
\end{equation}
From the definition of the generator
matrix $G_N=F^{\otimes n}$, it is shown in \cite{arikan_iti09} and \cite{gross_tc16}
that $G_{i,j} = 1$ if and only if $i \succeq j$, meaning the support of $\langle i-1\rangle_2$ contains the support of $\langle j-1\rangle_2$.
Let $i' \in \mathcal{\bar{A}}$ and $i \in \mathcal{A}$.
If $G_{i',i} = 1$, then we have $i' \succeq i$. From \cite{mori_icl09} and \cite{gross_tc16}, the binary
domination of $i'$ and $i$ indicates  bit channel upgradation: $W_N^{(i')} \succeq W_N^{(i)}$. But this contradicts
with the fact that bit channel $i' \in \mathcal{\bar{A}}$ is statistically degraded to all bit channels in the information
set $\mathcal{A}$. Therefore $G_{i',i}$ has to be zero.
\end{IEEEproof}

With $G_{\mathcal{\bar{A}A}}=0$, the systematic encoding (\ref{eq_xb_xbc}) can be simplified as
\begin{equation} \label{eq_xb_xbc3}
({x}_{\mathcal{A}},{x}_{\bar{\mathcal{A}}}) = ({u}_{\mathcal{A}}G_{\mathcal{AA}}, {u}_{\mathcal{A}}G_{\mathcal{A\bar{A}}}+{u}_{\bar{\mathcal{A}}}G_{\mathcal{\bar{A}\bar{A}}})
\end{equation}
Rewrite (\ref{eq_uacd}) as
\begin{equation} \label{eq_uacde}
u_{\mathcal{A}} = x_{\mathcal{A}}(G_{\mathcal{AA}})^{-1}
\end{equation}

As in \cite{arikan_icl11}, systematic encoding of polar
codes adopts the selection of  $\mathcal{B} = \mathcal{A}$. Efficient implementation
of systematic codes exists \cite{li_socc15}\cite{gross_tc16}. In \cite{li_socc15}, the efficient construction
of systematic polar codes resides on the fact that $G_{\mathcal{AA}}^{-1} = G_{\mathcal{AA}}$
in the binary field. In \cite{gross_tc16},
this condition is reformed as information set $\mathcal{A}$ being domination contiguous:
for $h,j \in \mathcal{A}$, and for $i \in \{1,2,...,N\}$, the following holds:
\begin{equation}\label{eq_domination}
\{h,j \in \mathcal{A} ~~\text{and}~~ (h-1) \succeq (i-1) \succeq (j-1)\} \implies i \in \mathcal{A}
\end{equation}
We now prove that the conditions in \cite{li_socc15} and \cite{gross_tc16} are essentially the same.


To prove the equivalence of the two efficiently encodable conditions in \cite{li_socc15} and \cite{gross_tc16},
the notations in \cite{gross_tc16} are introduced below. Let the information
bit channel set $\mathcal{A}=\{\alpha_i\}_{i=1}^{K}$ be an ascending
ordered set. Define a matrix $E$ (with a size of $K$ by $N$) as
\begin{eqnarray}\label{eq_e}
E=(E_{i,j})_{i=1,j=1}^{K,N}, ~~~\text{where~~~} E_{i,j} = \bigg\{
\begin{array}{cl}
1, & \text{if~~} j = \alpha_i \\
0, & \text{otherwise}
\end{array}
\end{eqnarray}
Let us first look at an example of this matrix $E$. Suppose $N=4$ and $\mathcal{A}=\{2,3,4\}$. Then  $E$ is
\begin{gather*}
E = \begin{bmatrix} 0 & 1 & 0 & 0
\\ 0 & 0 & 1 & 0
\\ 0 & 0 & 0 & 1 \end{bmatrix}
\end{gather*}
In \cite{gross_tc16}, with $\mathcal{A}$ being domination contiguous, the matrix $E \cdot F^{\otimes n} \cdot E^T$ is
proven to be an involution:
\begin{equation}\label{eq_involution}
\left(E\cdot F^{\otimes n} \cdot E^T\right) \left(E\cdot F^{\otimes n} \cdot E^T \right)= I
\end{equation}
where $I$ is the identity matrix. This is the basis for the efficient structure of systematic encoding for polar
codes in \cite{gross_tc16}. The following proposition shows that the condition in (\ref{eq_involution}) is essentially
the same as the efficient encoding condition in \cite{li_socc15}.
\begin{proposition}\label{proposition_equivalence}
The condition
$\left(E\cdot F^{\otimes n} \cdot E^T\right) \left(E\cdot F^{\otimes n} \cdot E^T \right)= I$
obtained from $\mathcal{A}$ being domination contiguous
is equivalent to $G_{\mathcal{AA}} = G_{\mathcal{AA}}^{-1}$.
\end{proposition}
\begin{IEEEproof}
First note that the generator matrix $G_N = F^{\otimes n}$. The operation $E\cdot F^{\otimes n}$ is then
$E\cdot F^{\otimes n} =E\cdot G_N $.
From the definition of $E$ in (\ref{eq_e}), the result of $E\cdot G_N$ is to take the rows in $\mathcal{A}$ from
the matrix $G_N$, denoted as $G_{\mathcal{A},:}$. Then $E\cdot F^{\otimes n} \cdot E^T = (G_{\mathcal{A},:}) \cdot E^T$.
Similarly, the operation of $G_{\mathcal{A},:} \cdot E^T$ is taking columns $\mathcal{A}$ of $G_{\mathcal{A},:}$, which
indicates that $(G_{\mathcal{A},:}) \cdot E^T =G_{\mathcal{AA}}$. Therefore, the condition $\left(E\cdot F^{\otimes n} \cdot E^T\right) \left(E\cdot F^{\otimes n} \cdot E^T \right)= I$ is equivalent to
$G_{\mathcal{AA}} \cdot G_{\mathcal{AA}} = I$.
\end{IEEEproof}

From (\ref{eq_xb_xbc3}) and (\ref{eq_uacde}) and that $G_{\mathcal{AA}}^{-1} = G_{\mathcal{AA}}$ \cite{li_socc15}, ${u}_{{\mathcal{A}}}$ can be solved directly without going through the structure related to ${u}_{\bar{\mathcal{A}}}$.
What's more important is that the non-systematic and the systematic encoding structure is essentially the same
because  $G_{\mathcal{AA}}^{-1} = G_{\mathcal{AA}}$.
Such systematic selection is considered as efficiently encodable. In the sequel, we discuss  pilot
selections which satisfy the efficiently encodable condition:
\begin{equation}\label{eq_efficient}
G_{\mathcal{AA}}^{-1} = G_{\mathcal{AA}}
\end{equation}

\section{Pilot Selection Schemes}\label{sec_pilot_selection}
The general pilot selection needs to meet  the channel estimation requirements.
In an OFDM system, channel estimation can be done by setting some of the sub-carriers to be pilots to account
for the frequency variation of the channel. In the mean time, wireless channels can be time varying. Therefore,
pilots in the time domain are also inserted. To make tradeoff between transmission efficiency and channel
estimation accuracy, pilots are often sparsely inserted in the frequency or the time domain \cite{3gpp_13}.
In this section, the transmission model of this paper  is first introduced and then pilot
selections are discussed.

\subsection{Transmission Model}
In this section, the transmission model is discussed. The encoded binary bits in $\mathbf{x}$
is transmitted through the underlying channel $W(y|x)$. Denote a matrix $X=\text{diag}\{{{x}_1^N}\}$ as
a diagonal matrix with elements taken from the codeword ${x}_1^N$. The received signal is then
\begin{equation}\label{eq_y}
{y}_1^N = {h}_1^N X + {z}_1^N
\end{equation}
where ${h}_1^N=(h_1,h_2,...,h_N)$ is the channel response for each coded symbol and ${z}_1^N$ is
the AWGN noise vector with each element having mean zero and variance $N_0/2$.
Assume there is no inter-symbol interference (ISI) in this model.
In this paper, the channel
${h}_1^N$ is assumed to follow the Rayleigh distribution with a Doppler shift $f_d$.
With Jake's spectrum, the time correlation of
the channel can be described by the first kind of 0-th order Bessel function:
\begin{equation}\label{eq_rhh}
R_{{hh}}(k) = J_0(2\pi f_d k T)
\end{equation}
where $R_{{hh}}$ is the autocorrelation function of the channel and $T$ is the symbol duration.
In the next subsections, pilot selections are discussed in order to estimate the channel ${h}_1^N$.

\subsection{Efficient Selection Criterion}\label{sec_efficient_selection}
Denote $\mathcal{P}_f$ as pilot positions in $\bar{\mathcal{A}}$ and $\mathcal{P}_i$
as the pilot positions in ${\mathcal{A}}$.
Then $[x_{\mathcal{P}_f},x_{\mathcal{P}_i}]$ are known pilot symbols.
The encoding of polar codes with pilot selections is equivalent to the following problem:
for each information vector $x_{\mathcal{A}}$ and
the construction conditions: $\{\mathcal{A},u_{\bar{\mathcal{A}}},\mathcal{P}_f$, $\mathcal{P}_i\}$,
how to calculate $u_{\mathcal{A}}$ in order to produce $x_{\bar{\mathcal{A}}}$?
One can immediately observe  that this problem has no solution since the linear equation behind (\ref{eq_uacde})
only needs a length $K$ vector $x_{\mathcal{A}}$. However, in the pilot selection case, there are additional
$|\mathcal{P}_f|$ known values in $x_{\bar{\mathcal{A}}}$. Note that the known pilots $x_{\mathcal{P}_i}$ are
imbedded in the information bits $x_\mathcal{A}$. The new encoding problem with pilots in $x_{\bar{\mathcal{A}}}$
is therefore less constrained. Fig.~\ref{fig_encoding_pilot} shows such an encoding problem with $N=8$ and $R=0.5$.
In Fig.~\ref{fig_encoding_pilot} two pilots are selected: symbol 3 and symbol 6. Symbol 3 is from the set
$\mathcal{\bar{A}}$ and symbol $6$ is in the information set $\mathcal{A}$. Therefore, there are
5 known values from the right-hand side while only 4 unknowns ($u_{\mathcal{A}}$) are
required in the original systematic encoding of polar codes.

\begin{figure}
{\par\centering
\resizebox*{3.0in}{!}{\includegraphics{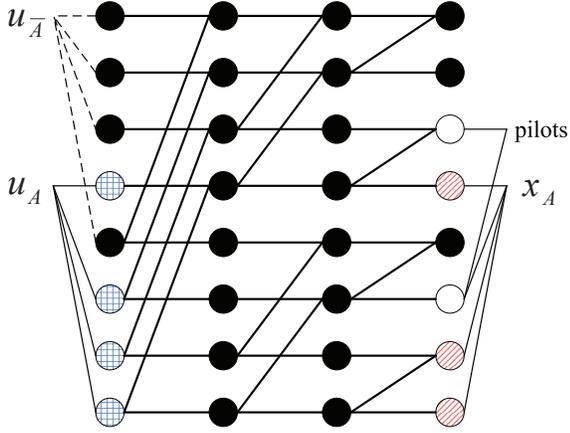}} \par}
\caption{The systematic encoding of polar codes with
pilot selections. The information set is $\mathcal{A}=\{4,6,7,8\}$. The pilots are selected as symbol 3 and 6.}
\label{fig_encoding_pilot}
\end{figure}

To make the encoding problem render a unique solution, one has to add more constraints.
Specifically, $|\mathcal{P}_f|$ constraints are needed. This means some of the frozen bits in $u_{\bar{\mathcal{A}}}$
can not be frozen anymore.
Let the union of the information set and the pilots in $\bar{\mathcal{A}}$ be $\mathcal{C}=\mathcal{A}~\cup~ \mathcal{P}_f$.
The encoding procedure can now be expressed as:
\begin{equation} \label{eq_xc_xcc}
({x}_{\mathcal{C}},{x}_{\bar{\mathcal{C}}}) = ({u}_{\mathcal{C}}G_{\mathcal{CC}}+{u}_{\bar{\mathcal{C}}}G_{\bar{\mathcal{C}}\mathcal{C}}, {u}_{\mathcal{C}}G_{\mathcal{C\bar{C}}}+{u}_{\bar{\mathcal{C}}}G_{\mathcal{\bar{C}\bar{C}}})
\end{equation}
One important note about this new encoding with pilot selection is that the set $\mathcal{C}$ is
no longer the information set as in the original encoding in (\ref{eq_xb_xbc_2}).
Instead, it includes elements in $\bar{\mathcal{A}}$ from the definition of $\mathcal{C}$.
Therefore, the {validity} of this new encoding needs to be first verified. Then the efficient
encodable mapping needs to be established.

The first condition of a valid mapping in (\ref{eq_sys_con_1}) is trivial:
$\mathcal{C}$ has the same elements as $\mathcal{C}$.
The second condition (\ref{eq_sys_con_2}) is that the mapping is one-to-one or $G_{\mathcal{CC}}$ is invertible.
This condition can also be easily verified as in \cite{arikan_icl11}: $G_{\mathcal{CC}}$ is lower triangular
with 1s at the diagonal and is therefore invertible.

Now comes to the efficient construction of this new selection in (\ref{eq_xc_xcc}). It is already
pointed out in Section \ref{sec_efficient_construction} that the efficient construction
relies on the fact that  $G_{\mathcal{CC}}^{-1} = G_{\mathcal{CC}}$. The following lemma shows that
this condition is met if $G_{\bar{\mathcal{C}}\mathcal{C}} = \mathbf{0}$.
\newtheorem{lemma}{Lemma}
\begin{lemma}\label{lemma_2}
Let $\mathcal{C} \subseteq \{1,2,...,N\}$. When $G_{\mathcal{\bar{C}C}} = \mathbf{0}$, then $G_{\mathcal{CC}}^{-1}=G_{\mathcal{CC}}$.
\end{lemma}
\begin{IEEEproof}
The proof is as in \cite{li_socc15}.
With $G_{\mathcal{\bar{C}C}}=\mathbf{0}$, $x_{\mathcal{C}}$ in (\ref{eq_xc_xcc}) can be written as
\begin{equation} \label{eq_xc}
x_{\mathcal{C}} = u_{\mathcal{C}}G_{\mathcal{CC}}
\end{equation}
which is equivalent as:
\begin{equation} \label{eq_uc}
u_{\mathcal{C}} = x_{\mathcal{C}}G_{\mathcal{CC}}^{-1}
\end{equation}
as $G_{\mathcal{CC}}$ is an invertible matrix (being a lower triangular matrix with ones at the diagonal).
In the mean time, from the encoding process ${x}_1^N = {u}_1^NG_N$ and $G_N^{-1}=G_N$ \cite{li_socc15},
we have
\begin{equation}\label{eq_u}
{u}_1^N = {x}_1^N G_N
\end{equation}
Decompose the vector ${u}_1^N$ in (\ref{eq_u}) as:
\begin{equation}\label{eq_uc_ucc}
(u_{\mathcal{C}},~~u_{\mathcal{\bar{C}}})=(x_{\mathcal{C}}G_{\mathcal{CC}},~~ x_{\mathcal{C}}G_{\mathcal{C\bar{C}}}+x_{\mathcal{\bar{C}}}G_{\mathcal{\bar{C}\bar{C}}})
\end{equation}
Since $G_{\mathcal{CC}}$ is an invertible matrix, (\ref{eq_uc}) and the first part of (\ref{eq_uc_ucc})
is equivalent as: $G_{\mathcal{CC}}$ = $G_{\mathcal{CC}}^{-1}$ in the binary field.
\end{IEEEproof}

From Lemma \ref{lemma_2}, the efficient construction condition can be checked from the submatrix $G_{\mathcal{\bar{C}C}}$.
Note that this condition from Lemma \ref{lemma_2}, unlike the condition in \cite{li_socc15} or \cite{gross_tc16}, does not
involve matrix inversion or matrix multiplication. Therefore $G_{\mathcal{\bar{C}C}} = {\mathbf{0}}$
is a simplified
working condition to check whether a selection $\mathcal{C}$ is efficiently encodable. In the rest of the paper,
$G_{\mathcal{\bar{C}C}} = {\mathbf{0}}$ is used to check the proposed pilot selection schemes
to determine whether they are efficiently encodable.

Compared with the submatrix $G_{\bar{\mathcal{A}}\mathcal{A}}$, $G_{\bar{\mathcal{C}}\mathcal{C}}$ has
$|\mathcal{P}_f|$ less rows but $|\mathcal{P}_f|$ more columns.
With $\mathcal{C}$ containing elements in $\mathcal{\bar{A}}$, $G_{\bar{\mathcal{C}}\mathcal{C}}$
is not guaranteed to be an all-zero matrix. In other words, the proof in \cite{li_socc15}
is not applicable for $G_{\bar{\mathcal{C}}\mathcal{C}}$. The set $\mathcal{C}$ is not necessarily
domination contiguous.

However, this efficient encodable problem is still promising due to the special or the
sparse nature of the generator matrix $G_N$. In the next two sections, we propose two pilot
selections which are efficiently encodable. Before going further, the following lemma is immediately
available.

\begin{lemma}\label{lemma_1}
Let $\mathcal{C} = \mathcal{A} ~\cup~ \mathcal{P}_f$ where $\mathcal{P}_f \subseteq \mathcal{\bar{A}}$.
Compared with $G_{\mathcal{\bar{A}A}}$, all increased columns
of $G_{\mathcal{\bar{C}C}}$ are from $G_{\mathcal{\bar{A}\bar{A}}}$.
\end{lemma}
\begin{IEEEproof}
The complementary set of $\mathcal{C}$ has less elements compared with
$\mathcal{\bar{A}}$: $\mathcal{\bar{C}} = \mathcal{\bar{A}} \backslash \mathcal{P}_f$ (Here $\backslash$ is the
excluding operation).
The submatrix $G_{\mathcal{\bar{C}C}}$ is
\begin{eqnarray}
G_{{\mathcal{\bar{C}C}}} &=& G_{(\mathcal{\bar{A}} \backslash \mathcal{P}_f)(\mathcal{{A}} \cup \mathcal{P}_f)} \\ \label{eq_ccc}
&=&[G_{(\mathcal{\bar{A}} \backslash \mathcal{P}_f)\mathcal{{A}}} ~~~ G_{(\mathcal{\bar{A}} \backslash \mathcal{P}_f)\mathcal{P}_f}]
\end{eqnarray}
The first part of (\ref{eq_ccc}): $G_{(\mathcal{\bar{A}} \backslash \mathcal{P}_f)\mathcal{{A}}}$, has the same
columns as $G_{\mathcal{\bar{A}A}}$. The second part of (\ref{eq_ccc}):
$G_{(\mathcal{\bar{A}} \backslash \mathcal{P}_f)\mathcal{P}_f}$ contains the additional columns of $G_{\mathcal{\bar{C}C}}$.
Since $\mathcal{\bar{A}} \backslash \mathcal{P}_f \subseteq \mathcal{\bar{A}}$ and $\mathcal{P}_f \subseteq \mathcal{\bar{A}}$,
the matrix $G_{(\mathcal{\bar{A}} \backslash \mathcal{P}_f)\mathcal{P}_f}$ is from $G_{\mathcal{\bar{A}\bar{A}}}$.
Therefore, compared with $G_{\mathcal{\bar{A}{A}}}$, the additional columns of $G_{\mathcal{\bar{C}C}}$
are from $G_{\mathcal{\bar{A}\bar{A}}}$.
\end{IEEEproof}

\subsection{Uneven Pilot Selection (UEPS)}\label{sec_case_one}
The matrix $G_{\mathcal{\bar{A}\bar{A}}}$ is an invertible matrix: it is a lower-triangular matrix
with ones at the diagonal in the binary field.
A detailed observation of $G_{\mathcal{\bar{A}\bar{A}}}$ reveals that some of the columns are all zeros except
the diagonal elements.
Denote $w(\cdot)$ as the Hamming weight of the inside argument.
The following set $\mathcal{S}$ is defined over $G_{\mathcal{\bar{A}\bar{A}}}$ as:
\begin{equation}\label{eq_S}
\mathcal{S} = \{j:~~  j \in \mathcal{\bar{A}} \text{~~and~~} w(G_{\mathcal{\bar{A}},j}) = 1 \}
\end{equation}
where $G_{\mathcal{\bar{A}},j}$ is the $j$th column of the submatrix $G_{\mathcal{\bar{A}},:}$. Remember that,
as in the proof of Proposition \ref{proposition_equivalence},
$G_{\mathcal{\bar{A}},:}$ is a submatrix formed by taking rows $\mathcal{\bar{A}}$ of $G_N$.
The following selection of pilots yields an efficiently encodable scheme.
\begin{proposition} \label{proposition_1}
Uneven Pilot Selection (UEPS): Let $\mathcal{P}_f \subseteq \mathcal{S}$ and the pilots in the set
$\mathcal{A}$ can be any desired selections. Then $\mathcal{C}=\mathcal{A} ~\cup~ \mathcal{P}_f$
yields $G_{\mathcal{\bar{C}}\mathcal{C}} = \mathbf{0}$.
\end{proposition}
\begin{IEEEproof}
Since $\mathcal{\bar{A}}\backslash\mathcal{P}_f \subseteq \mathcal{\bar{A}}$, the first part
of (\ref{eq_ccc}): $G_{(\mathcal{\bar{A}}\backslash\mathcal{P}_f)\mathcal{A}} = \mathbf{0}$ from
{{Proposition}} \ref{proposition_Gaca}.
Now consider the second part of (\ref{eq_ccc}).
The rows and columns of the matrix $G_{(\mathcal{\bar{A}} \backslash \mathcal{P}_f)\mathcal{P}_f}$
are from the set $\mathcal{\bar{A}}$. Since the matrix $G_{\mathcal{\bar{A}\bar{A}}}$ is a lower
triangular matrix with ones at the diagonal, the columns $\mathcal{S}$ of $G_{\mathcal{\bar{A}\bar{A}}}$
possess ones at the diagonal and are zeros elsewhere.
With $\mathcal{P}_f \in \mathcal{S}$, the non-zero elements surely only appear at the diagonal positions.
However $\mathcal{\bar{A}} \backslash \mathcal{P}_f$ excludes these diagonal positions.
Therefore $G_{(\mathcal{\bar{A}} \backslash \mathcal{P}_f)\mathcal{P}_f} = \mathbf{0}$.
According to Lemma \ref{lemma_1}, this means $G_{\mathcal{\bar{C}C}} = \mathbf{0}$.
\end{IEEEproof}

With  Lemma \ref{lemma_2}, the selection in Proposition \ref{proposition_1} meets
the efficiently encodable condition in (\ref{eq_efficient}) and is therefore an efficiently encodable selection.
Take $n=4$ and $R=0.5$ as an example. The encoding process selects the information set as $\mathcal{{A}}=\{8,10,11,12,13,14,15,16\}$ and the frozen set is $\mathcal{\bar{A}}=\{1,2,3,4,5,6,7,9\}$.
The submatrix $G_{\mathcal{\bar{A}\bar{A}}}$ is provided below:
\begin{gather*}
G_{\mathcal{\bar{A}\bar{A}}} = \begin{bmatrix} 1 & 0 & 0 & 0 & 0 & 0& 0& 0
\\ 1 & 1 & 0 & 0 & 0 & 0& 0& 0
\\ 1 & 0 & 1 & 0 & 0 & 0& 0& 0
\\ 1 & 1 & 1 & 1 & 0 & 0& 0& 0
\\ 1 & 0 & 0 & 0 & 1 & 0& 0& 0
\\ 1 & 1 & 0 & 0 & 1 & 1& 0& 0
\\ 1 & 0 & 1 & 0 & 1 & 0& 1& 0
\\ 1 & 0 & 0 & 0 & 0 & 0& 0& 1 \end{bmatrix}
\end{gather*}
Note that the last column (column eight) of $G_{\mathcal{\bar{A}\bar{A}}}$ is actually column nine of $G_{\mathcal{\bar{A}},:}$. Other columns of $G_{\mathcal{\bar{A}\bar{A}}}$ correspond to the same
columns of $G_{\mathcal{\bar{A}},:}$.
Then, from the submatrix $G_{\mathcal{\bar{A}\bar{A}}}$,
the set $\mathcal{S}$ can be
obtained as $\mathcal{S} = \{4,6,7,9\}$.  If $\mathcal{P}_f {{\subseteq}} \mathcal{S}$, then
$G_{\mathcal{\bar{C}}\mathcal{C}} = \mathbf{0}$.

Given the pilot selection as in Proposition \ref{proposition_1}, it is of interest to link
it with the domination contiguous condition in \cite{gross_tc16}. The following proposition
shows that the pilot selection in Proposition \ref{proposition_1} produces a domination contiguous set $\mathcal{C}$.
\begin{proposition}\label{proposition_1d}
The set $\mathcal{C} = \mathcal{A} ~\cup~ \mathcal{P}_f$  where $\mathcal{P}_f \subseteq \mathcal{S}$ is domination contiguous.
\end{proposition}
Please refer to Appendix \ref{appendix_1} for the proof of Proposition \ref{proposition_1d}.

However the pilots selected according to Proposition \ref{proposition_1} can not be evenly distributed among bit
channels $1$ to $N$. The pilot positions are dependent on the information set $\mathcal{A}$. For the same
block length and code rate, different channel conditions produce different information sets $\mathcal{A}$.
This makes the pilot selection inconsistent among channels and therefore can not be flexibly configured.
These drawbacks motivate
us to explore the structure of the encoding matrix $G_N$ to find controllable pilot selections which
are not dependent on the information set $\mathcal{A}$.



\subsection{Even Pilot Selection (EPS)}\label{sec_case_two}
Before the introduction of the pilot selection in this section, a new set $\mathcal{D}$ is defined as
 \begin{equation}\label{eq_D}
 \mathcal{D} =\{4k, ~~ 1 \le k \le N/4\}
 \end{equation}
The submatrix $G_{\mathcal{\bar{D}}\mathcal{D}}$ of $G_N$ is an all-zero matrix as stated in the following proposition.

\begin{proposition}\label{proposition_d}
With $\mathcal{D}$ defined in (\ref{eq_D}), the submatrix $G_{\mathcal{\bar{D}}\mathcal{D}}$ of
$G_N$ is an all-zero matrix: $G_{\mathcal{\bar{D}}\mathcal{D}} = \mathbf{0}$.
\end{proposition}
\begin{IEEEproof}
The generator matrix is $G_N=F^{\otimes n}$ where $F=\bigl(\begin{smallmatrix} 1&0 \\ 1&1\end{smallmatrix}\bigr)$.
The matrix $G_N$ can be decomposed as:
\begin{equation}\label{eq_GN}
G_N = F^{\otimes (n-2)} \otimes G_4
\end{equation}
Observe the matrix $G_{4}$:
\begin{equation}\label{eq_G3}
G_{4} =\begin{bmatrix}
       1 & 0 & 0 & 0   \\
       1 & 1 & 0 & 0   \\
       1 & 0 & 1 & 0    \\
       1 & 1 & 1 & 1
     \end{bmatrix}
\end{equation}
The fourth column of $G_4$ has one non-zero element at the fourth position. Now
extracting the submatrix of $G_N$ by selecting the columns specified by the
set $\mathcal{D}$ in (\ref{eq_D}). Denote $G_{:,\mathcal{D}}$ as such
a matrix.
From the expression of $G_N$ in (\ref{eq_GN}) and that only the fourth element of the fourth column
of $G_4$ is non-zero, it can be immediately concluded that the non-zero elements of $G_{:,\mathcal{D}}$
only appear in rows specified by the set $\mathcal{D}$. In other words, $G_{\mathcal{\bar{D}}\mathcal{D}} = \mathbf{0}$.
\end{IEEEproof}

For a given information set $\mathcal{A}$, let $\mathcal{D}_i = \mathcal{A} \cap \mathcal{D}$ and
$\mathcal{D}_f = \mathcal{\bar{A}} \cap \mathcal{D}$.
The following proposition states the second pilot selection which is efficiently encodable.


\begin{proposition}\label{proposition_p2}
Even Pilot Selection (EPS):
{{
Let $\mathcal{P}_i$ be the set containing any desired pilots in the information set $\mathcal{A}$,
and the pilots in the frozen set is: $\mathcal{P}_f = \mathcal{D}_f$.
}}
Then with the set $\mathcal{C} = \mathcal{A} ~\cup~ \mathcal{P}_f$, the submatrix $G_{\mathcal{\bar{C}C}} = \mathbf{0}$.
\end{proposition}

\begin{IEEEproof}
Since $\mathcal{P}_f = \mathcal{D}_f$ in this selection, equation (\ref{eq_ccc}) can be rewritten as
\begin{equation}\label{eq_ct}
G_{{\mathcal{\bar{C}C}}} =[G_{(\mathcal{\bar{A}} \backslash \mathcal{D}_f)\mathcal{{A}}} ~~~ G_{(\mathcal{\bar{A}} \backslash \mathcal{D}_f)\mathcal{D}_f}]
\end{equation}
With $\mathcal{\bar{A}} \backslash \mathcal{D}_f \in \mathcal{\bar{A}}$ and $G_{\mathcal{\bar{A}A}} = \mathbf{0}$,
the first part of (\ref{eq_ct}) is therefore an all-zero matrix:
$G_{(\mathcal{\bar{A}} \backslash \mathcal{D}_f)\mathcal{{A}}} = \mathbf{0}$.
The second part of (\ref{eq_ct}):
$G_{(\mathcal{\bar{A}} \backslash \mathcal{D}_f)\mathcal{D}_f}$ contains the additional columns of $G_{\mathcal{\bar{C}C}}$
compared with $G_{\mathcal{\bar{A}A}}$.
Since $\mathcal{\bar{A}} = \mathcal{D}_f + \mathcal{\bar{D}}_f$, then
$\mathcal{\bar{A}} \backslash \mathcal{D}_f = \mathcal{\bar{D}}_f$.
With
\begin{eqnarray}
\mathcal{{D}} = \mathcal{{D}}_i ~{{\cup}}~ \mathcal{{D}}_f \\
\mathcal{\bar{D}} = \mathcal{{\bar{D}}}_i ~{{\cup}}~ \mathcal{{\bar{D}}}_f
\end{eqnarray}
it can be immediately shown that
\begin{eqnarray}
\mathcal{D}_f \subseteq \mathcal{{D}}\\
\mathcal{\bar{D}}_f \subseteq \mathcal{\bar{D}}
\end{eqnarray}
Then the matrix $G_{(\mathcal{\bar{A}} \backslash \mathcal{D}_f)\mathcal{D}_f}$ is a submatrix from $G_{\mathcal{\bar{D}{D}}}$.
From Proposition \ref{proposition_p2}, $G_{\mathcal{\bar{D}{D}}} = \mathbf{0}$. Therefore
$G_{(\mathcal{\bar{A}} \backslash \mathcal{D}_f)\mathcal{D}_f} = \mathbf{0}$. Since the first part of (\ref{eq_ct}) is an
all-zero matrix and $G_{(\mathcal{\bar{A}} \backslash \mathcal{D}_f)\mathcal{D}_f} = \mathbf{0}$, the submatrix
$G_{\mathcal{\bar{C}C}}$ is therefore an all-zero matrix.
\end{IEEEproof}

Applying Lemma \ref{lemma_2}, the selection in Proposition \ref{proposition_p2} is also an efficiently
encodable selection because it meets the condition in (\ref{eq_efficient}). This selection does not
depend on the distribution of the set $\mathcal{A}$. The candidates of the pilots are always fixed ($\mathcal{D}$)
for a given block length $N$. The only requirement of this selection is to make sure all the elements in
$\mathcal{D}_f = \mathcal{D} ~\cap~ \mathcal{\bar{A}}$ are selected as pilots.

Fig.~\ref{fig_pilot_case_1_2}
shows the pilots selected for polar codes with $N=256$ and $R=0.5$. The underlying channel is the AWGN channel with
{{ an $E_b/N_0$ of $3$ dB (BPSK is applied here).}}
There are total 64 pilots selected for both UEPS and EPS. In EPS, the pilots from the information set is
$\mathcal{P}_i = \mathcal{D}_i$.
From Fig.~\ref{fig_pilot_case_1_2}, the pilots selected from Proposition \ref{proposition_1}
(UEPS) are not evenly distributed while pilots from Proposition \ref{proposition_p2} (EPS) is guaranteed
to be evenly distributed. For UEPS, if the {{number of}} elements of the set $\mathcal{S}$ is less than
the number of pilots to be selected, then pilots from the information set can be selected to
be evenly distributed. The pilots of UEPS in Fig.~\ref{fig_pilot_case_1_2} are selected in such a way.
However, even with this optimization, there are still gaps between pilots observed for UEPS in Fig.~\ref{fig_pilot_case_1_2}.
{{On}} the other hand,
the integer multiples of $4$ can be selected as pilots when needed in EPS. Therefore,
the channel estimation  performance of EPS in general should be better than {{that of}} UEPS.

\begin{figure}
{\par\centering
\resizebox*{3.0in}{!}{\includegraphics{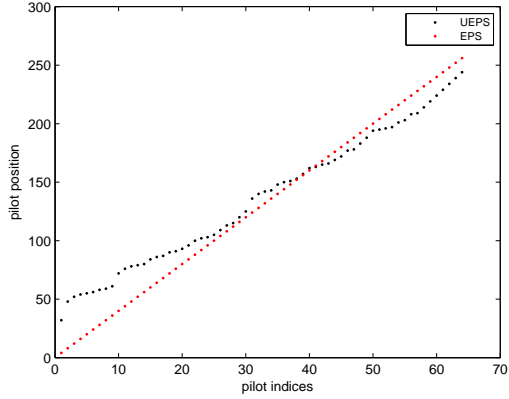}} \par}
\caption{The pilots selected according to Proposition \ref{proposition_1} (UEPS) and
Proposition \ref{proposition_p2} (EPS). The code length $N = 256$ and the code rate $R = 0.5$.
The information set $\mathcal{A}$ is selected based on an AWGN channel with {{an $E_b/N_0$}} of 3 dB.}
\label{fig_pilot_case_1_2}
\end{figure}
To conclude this section, we want to point out that the set $\mathcal{C} = \mathcal{A} ~\cup~ \mathcal{D}_f$
is domination contiguous.
\begin{proposition}\label{proposition_2d}
The  set $\mathcal{C} = \mathcal{A} \cup \mathcal{D}_f$ is domination contiguous.
\end{proposition}
Please refer to Appendix \ref{appendix_2} for the proof of Proposition \ref{proposition_2d}.

\subsection{Efficiency Comparison with Traditional Pilot Insertion}\label{sec_efficiency}
The two pilot selections in the previous two subsections are to use the coded symbols as pilots.
Traditional pilots are inserted into the existing codewords. For example, in LTE,
downlink pilots are inserted every four symbols in the time domain \cite{3gpp_13}.
Fig.~\ref{fig_pilot_selection_insertion} shows a comparison of EPS pilot selection and
the traditional pilot insertions.
\begin{figure}
\centering
\subfloat[EPS Pilot Selection]{\includegraphics[width=2.2in]{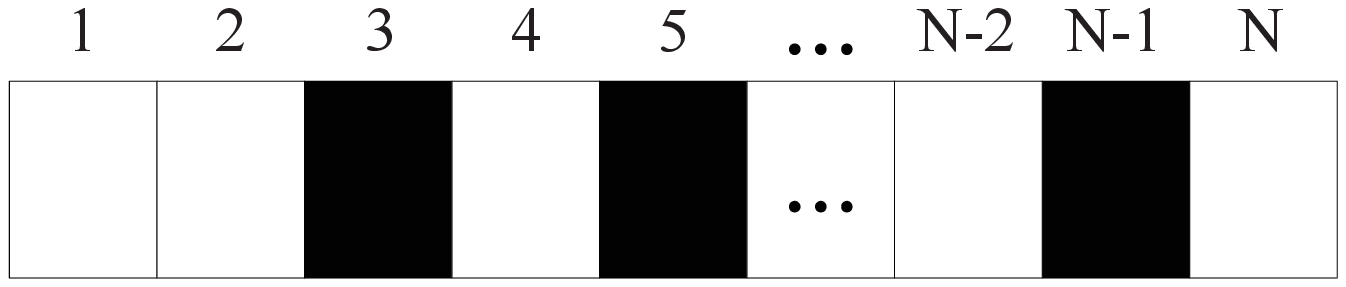}}\\
\subfloat[Traditional Pilot Insertion]{\includegraphics[width=2.2in]{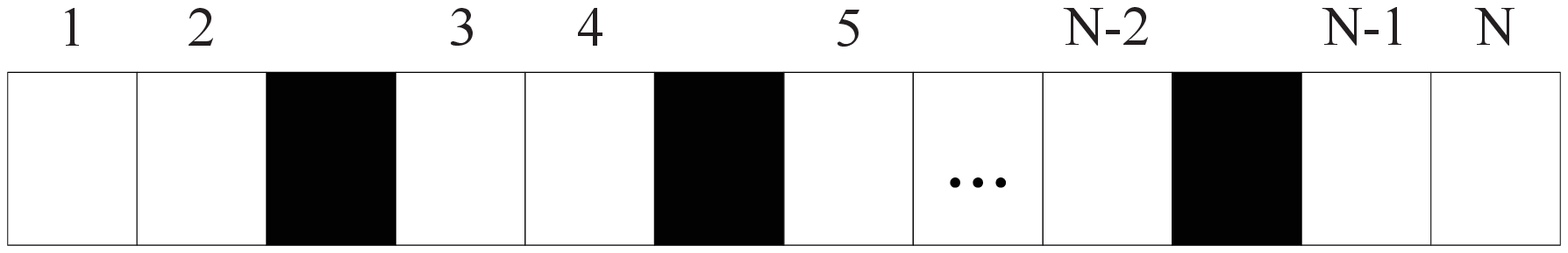}}
\caption{The numbers above the blocks are the symbol indices in one code block. The black
blocks are the pilots.
(a): EPS pilot selection where pilots are selected from the coded symbols.
(b): Traditional pilot insertions where pilots are additionally inserted among the coded symbols
}
\label{fig_pilot_selection_insertion}
\end{figure}

Suppose on average $K_P$ pilots are needed in one polar code block.
For UEPS and EPS pilot selections, denote $K_P = K_i + K_f$ where
$K_i$ is the number of pilots in the information set $\mathcal{A}$ and $K_f$ is the
number of pilots in the frozen set $\mathcal{\bar{A}}$.
The equivalent throughput of UEPS and EPS is therefore:
\begin{equation}\label{eq_Rp}
R_p = \frac{K-K_i}{N}
\end{equation}
The equivalent throughput of the traditional pilot insertions is:
\begin{equation}\label{eq_Rt}
R_t = \frac{K}{N+K_P}
\end{equation}
Assume the pilots are selected or inserted with an even spacing (which is the case for many practical
systems). Let $K_p = \alpha N$, where $\alpha$ is the ratio of pilots selected or inserted in one
code block. With the even spacing assumption, $K_i = R K_p$. With simple manipulations,
the ratio of $\gamma = R_p/R_t$ is:
\begin{equation}\label{eq_r_ratio}
\gamma = \frac{R_p}{R_t} = (1-\alpha)(1+\alpha)
\end{equation}

{\emph{\textbf{Remark}:}}
This ratio of $\gamma$ only works for EPS in Section \ref{sec_case_two} relative to the traditional
pilot insertion. UEPS in Section \ref{sec_case_one} does not have an evenly distributed
pilots and therefore violates the assumption of the even spacing between pilots. Also note
that the ratio $\gamma$ in (\ref{eq_r_ratio}) does not depend on the block length $N$
or the code rate $R$. It is only dependent on the ratio $\alpha$: the fraction of pilots
selected or inserted in one code block.

Fig.~\ref{fig_r_ratio} shows the ratio $\gamma$ with some values of $\alpha$. Pick a typical
value of $\alpha = 0.25$, the ratio $\gamma = 0.94$, meaning that EPS in Section \ref{sec_case_two}
only has a slightly smaller throughput compared with the traditional pilot insertion. However, as
analyzed in the next subsection and from the simulation results in Section \ref{sec_simulation}, the
performance of EPS is much better than {{that of}} the traditional pilot insertion.
{{Furthermore, the small throughput loss of UEPS and EPS can be overcome by initially setting
a larger code rate than that of the traditional pilot insertion. In Section \ref{sec_simulation}, with
the same throughput between EPS and the traditional pilot insertion, EPS still has a very clear
advantage in terms of the error performance.}}

\begin{figure}
{\par\centering
\resizebox*{3.0in}{!}{\includegraphics{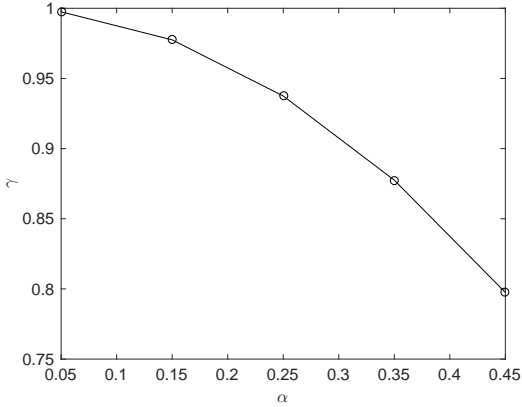}} \par}
\caption{The relative throughput $\gamma = R_p / R_t$ between the pilot selection and the traditional pilot insertion with $K_p = \alpha N$.}
\label{fig_r_ratio}
\end{figure}

\subsection{Decoding Processing with Pilot Selections}\label{sec_decoding}
The error performance of the pilot selections is the main motivation of this paper.
Bear in mind that, the traditional inserted pilots only facilitate the channel estimation.
The proposed pilot selections UEPS and EPS not only serve the purpose of channel
estimation, but also help in the decoding process. Fig.~\ref{fig_decoding_processing} shows the log likelihood ratio (LLR) values fed from the left-hand side and the right-hand side of the decoding graph. The parameters
of the polar code in Fig.~\ref{fig_decoding_processing} is the same as that in Fig.~\ref{fig_encoding_pilot} where $N=8$, $\mathcal{A}=\{4,6,7,8\}$ and two symbols, $x_3$ and $x_6$, are selected as pilots. Assume pilot symbols are all zeros. Then the LLR values corresponding to the
pilot symbols can be set to infinity in the decoding process.
\begin{figure}
{\par\centering
\resizebox*{3.0in}{!}{\includegraphics{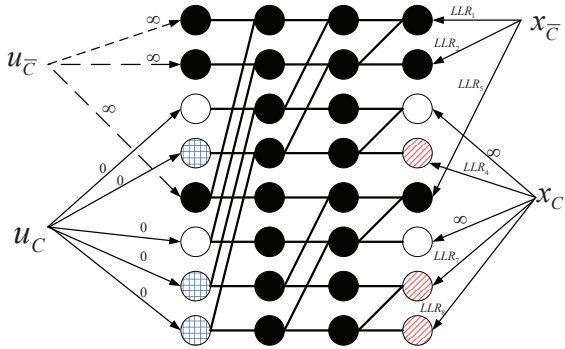}} \par}
\caption{The decoding processing with pilots selected from coded symbols.}
\label{fig_decoding_processing}
\end{figure}

Polar code decoding, be it the original {{successive}} cancellation (SC) decoding or the belief
propagation (BP) decoding, can be unified in the BP decoding frame \cite{urbanke_isit09}.
Also bear in mind that the generator matrix $G_N= F^{\otimes n}$ is an involution: $G_N  = G_N^{-1}$.
The involution property of the matrix $G_N$ can be interpreted as the following: a LLR value
fed from the left-hand side propagates through the same nodes and edges as that fed from the same position from the
right-hand side and therefore has the same contribution to the decoding performance.
For a given block length $N$ and code rate $R$, the decoding performance of polar codes can now be
characterized by the initial conditions set
\begin{eqnarray} \nonumber
\mathcal{L} &=& \{LLR(y_{\mathcal{A}}), LLR(y_{\mathcal{\bar{A}}}), LLR(u_{\mathcal{\bar{A}}}), LLR(u_{\mathcal{A}})\} \\ \nonumber
&=&\{LLR(y_{\mathcal{A}}), LLR(y_{\mathcal{\bar{A}}}), \\\label{eq_b}
& &LLR(u_{\mathcal{\bar{A}}}) = \infty_1^{N-K}, LLR(u_{\mathcal{A}}) = 0_1^K\}
\end{eqnarray}
where the vector $y_1^N$ contains the received samples from the channel and
$LLR(\cdot)$ is the LLR values of the inside {{arguments}}.
Apparently,  when no symbols are selected as pilots, there are $N-K$ frozen
bits ($LLR(u_{\mathcal{\bar{A}}}) = \infty_1^{N-K}$) and the LLRs corresponding to the information bits
are $LLR(u_{\mathcal{A}}) = 0_1^K$.

When analyzing the effect of pilot selections to the
decoding performance, the pilots in the frozen set
$\mathcal{\bar{A}}$ and pilots in $\mathcal{A}$ have to be differentiated. First let us look at the pilots in
the frozen set $\mathcal{\bar{A}}$. Without loss
of generality, take EPS in Section \ref{sec_case_two} as an example. For any $j \in \mathcal{D}_f$, a corresponding $u_j$
in the left-hand side is taken from the frozen set $\mathcal{\bar{A}}$, which results in a zero LLR value fed from the left-hand side to the decoding graph. This conversion of $u_j$ from the frozen bit to a non-frozen bit is a basic
requirement of the new encoding process in Section \ref{sec_efficient_selection}. The symbol $x_3$
in Fig.~\ref{fig_decoding_processing} is such an example: in the decoding process, the LLR of $y_3$ (the
received sample of $x_3$)
is fixed as infinity while the LLR of $u_3$ is 0.
The initial condition set is now:
\begin{eqnarray}\nonumber
\mathcal{L}_f &=& \{(LLR(y_{\mathcal{A}}), LLR(y_{\mathcal{\bar{A}}/\mathcal{P}_f}), LLR(y_{\mathcal{P}_f}) = \infty_1^{K_f}, \\\nonumber
 &~& LLR(u_{\mathcal{\bar{A}}/\mathcal{P}_f}) = \infty_1^{(N-K-Kf)}, LLR(u_{\mathcal{P}_f})=0_1^{K_f}, \\ \label{eq_bf}
 &~& LLR(u_{\mathcal{A}}) = 0_1^K\}
\end{eqnarray}
Comparing the set $\mathcal{L}_f$ with the set $\mathcal{L}$ in (\ref{eq_b}), the infinite
LLR values feeding into the decoding diagram is the same: there are additional $K_f$
infinite LLR values from the right-hand side but $K_f$ less infinite LLR values from
the left-hand side. Therefore, the decoding error performance of polar codes given the input
set $\mathcal{L}$ and $\mathcal{L}_f$ should be on the same level.
In this case, we consider $\mathcal{L}$ and $\mathcal{L}_f$
as equivalent.


When a symbol $ i \in \mathcal{A}$ is selected as a pilot, the decoding performance is improved
compared when symbol $i$ is a normal information bit. Since $\mathcal{L}$ and $\mathcal{L}_f$
are equivalent, here we only select $ i \in \mathcal{A}$ as pilots. The new initial set is
\begin{eqnarray}\nonumber
\mathcal{L}_i=\{LLR(y_{\mathcal{\bar{A}}}), LLR(y_{\mathcal{{A}}/\mathcal{P}_i}), LLR(y_{\mathcal{P}_i}) = \infty_1^{K_i}, \\ \label{eq_bi}
 LLR(u_{\mathcal{\bar{A}}}) = \infty_1^{(N-K)}, LLR(u_{\mathcal{A}})=0_1^{K}\}
\end{eqnarray}
Note that the initial $u_{\mathcal{P}_i} = 0_1^{K_i}$ is merged with $u_{\mathcal{A}\backslash\mathcal{P}_i}=0_1^{K-K_i}$
as $u_{\mathcal{A}} = 0_1^K$ in (\ref{eq_bi}).
The initial condition set $\mathcal{L}_i$ with $K_i$ pilots selected from $\mathcal{A}$ has $K_i$ additional
infinite LLR values compared with the set $\mathcal{L}$ in (\ref{eq_b}). These are stronger (or absolutely definite)
initial values which greatly benefit the decoding process. In this sense, the decoding performance
of the EPS scheme with at least one pilot from the information set should be better than
the traditional pilot insertion with the same number of pilots inserted. The simulation results in
Section \ref{sec_simulation} verified the analysis in this section.

\subsection{Channel Estimation Performance}\label{sec_mse}
The channel estimation performance in terms of mean square error (MSE) is analyzed in this section.
Two estimators are used in this paper: least square (LS) and minimum mean square error (MMSE) estimators.
Then linear interpolation is used to estimate the channel response at non-pilot positions.
Let the set $\mathcal{P} = \mathcal{P}_f ~{{\cup}}~ \mathcal{P}_i$ be the set containing all the pilot positions.
For the LS estimator, the estimation of the channel response at the pilot positions are \cite{gershman_tsp06}
\cite{cho_10}
\begin{equation}
\tilde{{h}}_{\mathcal{P}} = {y}_{\mathcal{P}}X_{\mathcal{PP}}^{-1}
\end{equation}
The MSE of the LS estimator is well established to be:
\begin{equation}
\text{MSE}_{\text{LS}} = \frac{1}{R{E_b/N_0}}
\end{equation}
While for the  MMSE estimation, the estimation of the channel response at the pilot positions are:
\begin{equation}
\hat{{h}}_{\mathcal{P}} = R_{{h}_{\mathcal{P}}{\tilde{h}}_{\mathcal{P}}}
\left(R_{{h}_{\mathcal{P}}{h}_{\mathcal{P}}}+\frac{1}{RE_b/N_0}I\right)^{-1}\tilde{{h}}_{\mathcal{P}}
\end{equation}
where the matrix $R_{\mathbf{ab}}$ is the cross correlation of the vector $\mathbf{a}$ and $\mathbf{b}$:
$R_{\mathbf{ab}} = \E\{\mathbf{a}\mathbf{b}^H\}$ and $I$ is the identity matrix.
Without simplifications and approximations, the MSE of the estimation ${\hat{h}_1^N}$ does not in general
yield a closed form expression. However, as shown in \cite{gershman_tsp06}
\cite{cho_10}, MMSE performs better than LS in low $E_b/N_0$ regions. Numerical results of the MSE
of EPS and UEPS with MMSE and LS are reported.

The pilot selection UEPS or EPS should have the same channel estimation performance as the traditional
pilot insertion given that pilots are inserted in the same positions as UEPS or EPS. However, UEPS in
general should have worse channel estimation performance than EPS due to the uneven nature of its pilots.
In the next section, the MSE performance of these schemes are compared.

\emph{\bf{Remark}}: When the channel responses $h_1^N$ are highly correlated (for example, in a static or a slowly moving
environment), the pilots {{$\mathcal{P}_i$}} (from the information set) of
EPS (or UEPS) can be made more sparse to increase
the throughput, while the pilots {{$\mathcal{P}_f$}} can still be
{{$\mathcal{D}_f$}} (or $\mathcal{S}$ in (\ref{eq_S}))
(these pilots do not affect the overall throughput).
In a fast-moving environment (with large Doppler frequencies),
although $\mathcal{P}_f=\mathcal{D}_f$ (or $\mathcal{S}$), $\mathcal{P}_i$ can be made as dense as needed.
Actually, one pilot every four symbols of EPS is already a very dense selection which should be able to meet requirements of most applications.

\section{Numerical Results}\label{sec_simulation}
In this section, the pilot selections in Section \ref{sec_pilot_selection} are simulated.  The channel is assumed to be the Rayleigh fading channel. Two channel estimators are compared: Least Square (LS) and
Minimum Mean Square Error (MMSE). Linear interpolation is used to estimate the
channel at non-pilot positions. The polar code simulated has  block length $N = 256$.
The encoded symbols are modulated with the BPSK scheme.
The SC decoding is applied in the decoding process.
The following wireless scenario is selected
as a test case: the carrier frequency is $900$ MHz and the symbol rate is $256$ Ksps.
Two Doppler frequencies $f_d = 10$ Hz and $f_d = 50$ Hz are tested in this section,
corresponding to {{two velocities}} of $12$ km/h and $60$ km/h, respectively.
The information set $\mathcal{A}$ is selected from the Tal-Vardy algorithm
in \cite{vardy_it13} with
an $E_b/N_0$ of $3$ dB (a further increase of the construction $E_b/N_0$ does not improve the error performance).

The MSE of the estimators is compared in Fig.~\ref{fig_mse} where the code rate is $R=0.5$.
As expected, for EPS, the LS estimator is not as good as the MMSE estimator. With the MMSE
estimator, the EPS scheme outperforms the UEPS scheme, also as expected
from the discussions in Section \ref{sec_mse}: the pilots in UEPS are not
evenly distributed as EPS. The MSE performance of UEPS is almost the same
as {{that of}} the EPS scheme with the LS estimator. In the following results, the
decoding performance echoes this observation.

The frame-error-rate (FER) performance of the EPS
in Section \ref{sec_case_two} is shown in Fig.~\ref{fig_fer_case2}.
The pilots selected are: $\mathcal{P}_f = \mathcal{D}_f$ and $\mathcal{P}_i = \mathcal{D}_i$.
The initial code rate of the polar code is $R=0.5$.
Remember that $\mathcal{D}_f$ and $\mathcal{D}_i$ contains elements (multiples of four)
in the frozen and information set, respectively.
From Fig.~\ref{fig_fer_case2},
it can be seen that the MMSE method with $f_d = 10$ Hz has a better FER performance than that of the LS with the
same Doppler shift. When $f_d$ goes up to $50$ Hz, the FER performance of MMSE and LS is worse than the corresponding
performance at $f_d = 10$ Hz.

The performance of the pilot selection UEPS in Section \ref{sec_case_one}
is compared with that of EPS in Fig.~\ref{fig_fer_case_one_two}. Note that pilots selected for UEPS are done
in two steps. First, the pilots in (\ref{eq_S}) are selected from the frozen set $\mathcal{\bar{A}}$.
Then, the remaining pilots are selected from the information set $\mathcal{A}$.
To achieve the best interpolation results, the pilots of UEPS in $\mathcal{A}$ are evenly selected.
Both EPS and UEPS have 64 pilots with 40 of them selected from the information set.
The Doppler shift in the simulation of
Fig.~\ref{fig_fer_case_one_two} is $f_d = 50$ Hz. The EPS with MMSE is better than UEPS with MMSE: at the FER of
$10^{-3}$, UEPS with MMSE requires 2 dB more than EPS with MMSE. For UEPS and EPS with LS, similar
phenomenon is observed. The superior performance of EPS pilot selection is due to the even distribution
of pilots rather than the unevenly distributed pilots in UEPS.

The two efficient pilot selections UEPS and EPS are compared with the traditional pilot
insertion in Fig.~\ref{fig_fer_case_one_two_traditional} where the Doppler shift is also $f_d=50$ Hz.
For the pilot insertion scheme, pilots are evenly inserted: one pilot is inserted every four coded symbols.
The polar code used for the pilot insertion has a block length $N=256$  and an initial code rate $R=0.5$.
The number of pilots in these three schemes are the same: total $64$ pilots are employed.
{{
According to the analysis in Section \ref{sec_efficiency}, the overall throughput of the traditional pilot insertion
is $R_t=128/(256+64)=0.4$. To maintain the same throughput of UEPS and EPS as that of the traditional pilot
insertion, the initial code rate of the polar code is adjusted as $R=147/256 = 0.574$. Among the $64$ pilot
symbols, $45$ of them are from $\mathcal{P}_i$, resulting in a final throughput of $R_p = (147-45)/256 = 0.4$
for both UEPS and EPS schemes.
}}
It can be seen from Fig.~\ref{fig_fer_case_one_two_traditional} that the UEPS and the traditional pilot
insertion has almost the same FER performance when MMSE is applied while EPS has a better FER performance.
Thus the 2 dB advantage of EPS over UEPS at FER $10^{-3}$ applies to the traditional pilot selection.
However, this advantage of EPS over the traditional pilot insertion comes from the fact that all pilots
inserted only serve as the channel estimation elements. For EPS pilot selection, the pilots also
facilitate the decoding of polar codes as discussed in Section \ref{sec_decoding}.


\begin{figure}
{\par\centering
\resizebox*{3.0in}{!}{\includegraphics{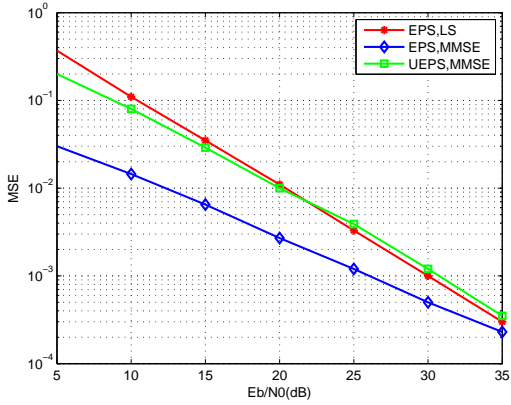}} \par}
\caption{The MSE performance of UEPS and EPS with MMSE and LS estimators.
The channel is the Rayleigh fading
channel with a Doppler frequency  $f_d=50$ Hz. In each code block of $N=256$, $64$ pilots
are selected.}
\label{fig_mse}
\end{figure}

\begin{figure}
{\par\centering
\resizebox*{3.0in}{!}{\includegraphics{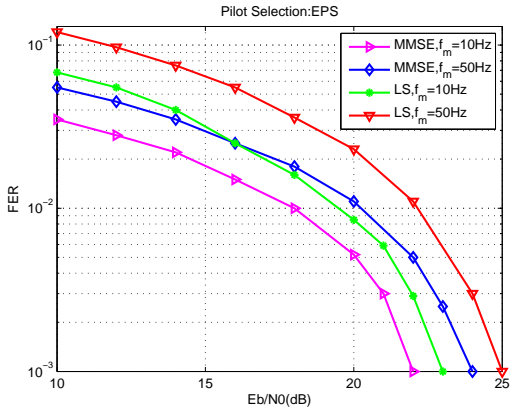}} \par}
\caption{The FER performance of channel estimation with systematic polar codes over the Rayleigh fading
channel with Doppler frequencies $f_d = 10$ Hz and $f_d=50$ Hz. The pilots are selected according to
the efficient pilot selection (EPS) in Section \ref{sec_case_two}. The polar code has the block length
$N=256$ and the code rate $R=0.5$.}
\label{fig_fer_case2}
\end{figure}

\begin{figure}
{\par\centering
\resizebox*{3.0in}{!}{\includegraphics{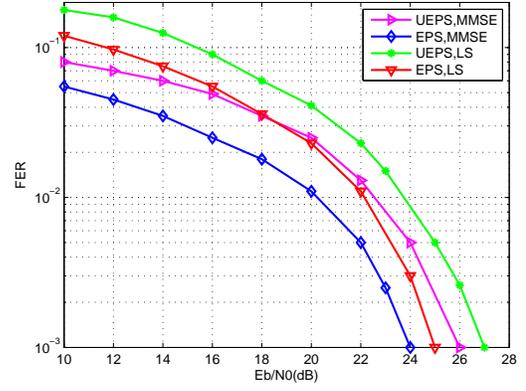}} \par}
\caption{The FER performance comparison of the pilot selections of UEPS (Section \ref{sec_case_one})
and EPS (Section \ref{sec_case_two}) in the Rayleigh fading channel with a Doppler spread $f_d=50$ Hz.
The polar code has the block length $N=256$ and the code rate $R=0.5$.}
\label{fig_fer_case_one_two}
\end{figure}

\begin{figure}
{\par\centering
\resizebox*{3.0in}{!}{\includegraphics{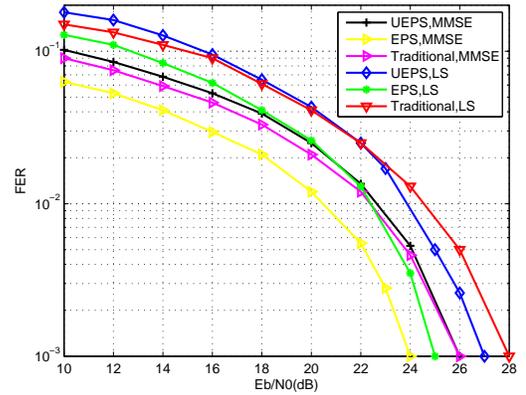}} \par}
\caption{The PER performance comparison of the pilot selections (UEPS and EPS) with
the traditional pilot insertion in a Rayleigh fading channel with a Doppler spread $f_d=50$ Hz.
{{For all schemes, the polar code has the block length $N=256$, and the number of pilots is 64.
For the traditional pilot insertion,  the code rate is $R=0.5$ and the overall throughput is $R_t=0.4$.
For UEPS and EPS, the initial code rate is $R=0.574$. With 45 pilots selected from $\mathcal{P}_i$, the
overall throughput for UEPS and EPS is $R_p=0.4$.
}}
}
\label{fig_fer_case_one_two_traditional}
\end{figure}

\section{Conclusion}\label{sec_con}
In this paper, two pilot selection schemes, uneven pilot selection (UEPS) and even pilot selection (EPS)
are studied for polar codes in fading channels. By
selecting coded symbols as pilots, instead of inserting pilots, the decoding performance
of polar codes {{is}} greatly improved. Considering the unsatisfactory performance of polar
codes in the finite domain, the proposed pilot selection scheme EPS can
be employed in practical systems for channel estimation or tracking. Simulation results
show that the proposed EPS scheme outperforms both the UEPS scheme and the traditional pilot
insertion scheme and therefore is a promising option to improve the polar code performance
in wireless communications.

\appendices
\section{Proof of Proposition \ref{proposition_1d}} \label{appendix_1}
Let $h, j \in \mathcal{C} = \mathcal{A} \cup \mathcal{P}_f$.
\begin{itemize}
\item If $h,j \in \mathcal{A}$, then any $i$ which satisfies $h-1 \succeq i-1$ and $i-1 \succeq j-1$ must
be in $\mathcal{A}$ since $\mathcal{A}$ is domination contiguous \cite{gross_tc16}. In this case, $i \in \mathcal{A} \subseteq \mathcal{C}$.
\item If $h \in \mathcal{A}$ and $j \in \mathcal{P}_f$, when $i \in \mathcal{A}$ satisfying $h-1 \succeq i-1$ and $i-1 \succeq j-1$, then  $i \in \mathcal{A} \subseteq \mathcal{C}$. When $i \in \mathcal{\bar{A}}$ satisfying $h-1 \succeq i-1$ and $i-1 \succeq j-1$, then from the selection of $\mathcal{P}_f \subseteq \mathcal{S}$ and the definition of $\mathcal{S}$ in
    (\ref{eq_S}), it can be concluded that there is no other elements in $\mathcal{\bar{A}}$ which satisfy $i-1 \succeq j-1$
    except $i=j$. Therefore $i=j \in \mathcal{P}_f \subset \mathcal{C}$.
\item If $h,j \in \mathcal{P}_f$, there is no  $i \in \mathcal{A}$ which meets $h-1 \succeq i-1$  since bit channel $h$ is worse than bit channel $i$ and therefore $h-1 \succeq i-1$ can never occur. If $i \in \mathcal{\bar{A}}$, then $i$ has to
    be the same as $j \in \mathcal{P}_f$ in order to have $i-1 \succeq j-1$.
\item If $h \in \mathcal{P}_f$ and $j \in \mathcal{A}$, to have $h-1 \succeq i-1$, then $i$ has to be from $\mathcal{\bar{A}}$.
To have $i-1 \succeq j-1$, $i$ has to be from $\mathcal{A}$. There is no $i$ to satisfy $h-1 \succeq i-1$ and $i-1 \succeq j-1$.
\end{itemize}
From the above analysis, for all $h, j \in \mathcal{C}$,  the $i$s which meet $h-1 \succeq i-1$ and $i-1 \succeq j-1$
are also in $\mathcal{C}$. This concludes that $\mathcal{C}$ is domination contiguous.

\section{Proof of Proposition \ref{proposition_2d}}\label{appendix_2}
The proof of this proposition is similar to Appendix \ref{appendix_1}.
Let $h, j \in \mathcal{C} = \mathcal{A} \cup \mathcal{D}_f$.
\begin{itemize}
\item If $h,j \in \mathcal{A}$, then any $i$ which satisfies $h-1 \succeq i-1$ and $i-1 \succeq j-1$ must
be in $\mathcal{A}$ since $\mathcal{A}$ is domination contiguous \cite{gross_tc16}. In this case, $i \in \mathcal{A} \subseteq \mathcal{C}$.
\item If $h \in \mathcal{A}$ and $j \in \mathcal{D}_f$, when $i \in \mathcal{A}$ satisfying $h-1 \succeq i-1$ and $i-1 \succeq j-1$, then  $i \in \mathcal{A} \subseteq \mathcal{C}$. When $i \in \mathcal{\bar{A}}$ satisfying $h-1 \succeq i-1$ and $i-1 \succeq j-1$, then the binary expansion of $i-1$ can be expressed $\langle i_1,i_2,...,1,1\rangle$. The last two bits are ones since $j$ is a multiple of 4 and $i-1 \succeq j-1$. The number $i$ with $\langle i-1 \rangle_2 = \langle i_1,i_2,...,1,1 \rangle$ is again a multiple of 4. Therefore $i \in \mathcal{D}_f$.
\item If $h,j \in \mathcal{D}_f$, there is no  $i \in \mathcal{A}$ which meets $h-1 \succeq i-1$
since bit channel $h$ is worse than bit channel $i$ and therefore $h-1 \succeq i-1$ can never occur. If $i \in \mathcal{\bar{A}}$, then $i$ has to be a multiple of 4 in order to have $i-1 \succeq j-1$. In this case $i \in \mathcal{D}_f$.
\item If $h \in \mathcal{D}_f$ and $j \in \mathcal{A}$, to have $h-1 \succeq i-1$, then $i$ has to be from $\mathcal{\bar{A}}$.
To have $i-1 \succeq j-1$, $i$ has to be from $\mathcal{A}$. There is no $i$ to satisfy $h-1 \succeq i-1$ and $i-1 \succeq j-1$.
\end{itemize}
From the above analysis, for all $h, j \in \mathcal{C}$,  the $i$s which meet $h-1 \succeq i-1$ and $i-1 \succeq j-1$
are also in $\mathcal{C}$. This concludes that $\mathcal{C}$ is domination contiguous.

\bibliography{ref_polar}

\begin{thebibliography}{10}
\providecommand{\url}[1]{#1}
\csname url@samestyle\endcsname
\providecommand{\newblock}{\relax}
\providecommand{\bibinfo}[2]{#2}
\providecommand{\BIBentrySTDinterwordspacing}{\spaceskip=0pt\relax}
\providecommand{\BIBentryALTinterwordstretchfactor}{4}
\providecommand{\BIBentryALTinterwordspacing}{\spaceskip=\fontdimen2\font plus
\BIBentryALTinterwordstretchfactor\fontdimen3\font minus
  \fontdimen4\font\relax}
\providecommand{\BIBforeignlanguage}[2]{{%
\expandafter\ifx\csname l@#1\endcsname\relax
\typeout{** WARNING: IEEEtran.bst: No hyphenation pattern has been}%
\typeout{** loaded for the language `#1'. Using the pattern for}%
\typeout{** the default language instead.}%
\else
\language=\csname l@#1\endcsname
\fi
#2}}
\providecommand{\BIBdecl}{\relax}
\BIBdecl

\bibitem{arikan_iti09}
E.~Arikan, ``{Channel Polarization: A Method for Constructing
  Capacity-Achieving Codes for Symmetric Binary-Input Memoryless Channels},''
  \emph{IEEE Transactions on Information Theory}, vol.~55, no.~7, pp.
  3051--3073, 2009.

\bibitem{mori_icl09}
R.~Mori and T.~Tanaka, ``{Performance of polar codes with the construction
  using density evolution},'' \emph{IEEE Communications Letters}, vol.~13,
  no.~7, pp. 519--521, Jul. 2009.

\bibitem{vardy_it13}
I.~Tal and A.~Vardy, ``{How to Construct Polar Codes},'' \emph{Information
  Theory, IEEE Transactions on}, vol.~59, no.~10, pp. 6562--6582, Oct 2013.

\bibitem{trifonov_itc12}
P.~Trifonov, ``{Efficient Design and Decoding of Polar Codes},'' \emph{IEEE
  Transactions on Communications}, vol.~60, no.~11, pp. 3221--3227, November
  2012.

\bibitem{wu_icl14}
D.~Wu, Y.~Li, and Y.~Sun, ``{Construction and Block Error Rate Analysis of
  Polar Codes Over {AWGN} Channel Based on {Gauss}ian Approximation},''
  \emph{IEEE Communications Letters}, vol.~18, no.~7, pp. 1099--1102, Jul.
  2014.

\bibitem{niu_itc13}
K.~Chen, K.~Niu, and J.~Lin, ``{Improved Successive Cancellation Decoding of
  Polar Codes},'' \emph{IEEE Transactions on Communications}, vol.~61, no.~8,
  pp. 3100--3107, August 2013.

\bibitem{vardy_it15}
I.~Tal and A.~Vardy, ``{List Decoding of Polar Codes},'' \emph{Information
  Theory, IEEE Transactions on}, vol.~61, no.~5, pp. 2213--2226, May 2015.

\bibitem{arikan_icl08}
E.~Arikan, ``{A Performance Comparison of Polar Codes and Reed-Muller codes},''
  \emph{IEEE Communications Letters}, vol.~12, no.~6, pp. 447--449, 2008.

\bibitem{urbanke_isit09}
N.~Hussami, S.~Korada, and R.~Urbanke, ``{Performance of Polar Codes for
  Channel and Source Coding},'' in \emph{IEEE International Symposium on
  Information Theory (ISIT)}, June 2009, pp. 1488--1492.

\bibitem{guo_isit14}
J.~Guo, M.~Qin, A.~G. i~Fabregas, and P.~H. Siegel, ``{Enhanced Belief
  Propagation Decoding of Polar Codes through Concatenation},'' in \emph{2014
  IEEE International Symposium on Information Theory Proceedings (ISIT)}, 2014,
  pp. 2987 -- 2991.

\bibitem{arikan_icl11}
E.~Arikan, ``{Systematic Polar Coding},'' \emph{IEEE Communications Letters},
  vol.~15, no.~8, pp. 860--862, August 2011.

\bibitem{li_tvt00}
Y.~Li, ``Pilot-symbol-aided channel estimation for ofdm in wireless systems,''
  \emph{IEEE Transactions on Vehicular Technology}, vol.~49, no.~4, pp.
  1207--1215, Jul 2000.

\bibitem{mengali_tsp01}
M.~Morelli and U.~Mengali, ``{A Comparison of Pilot-Aided Channel Estimation
  Methods for OFDM Systems},'' \emph{IEEE Transactions on Signal Processing},
  vol.~49, no.~12, pp. 3065--3073, December 2001.

\bibitem{li_twc02}
Y.~Li, ``Simplified channel estimation for ofdm systems with multiple transmit
  antennas,'' \emph{IEEE Transactions on Wireless Communications}, vol.~1,
  no.~1, pp. 67--75, Jan 2002.

\bibitem{gibson_twc05}
K.~J. Kim, J.~Yue, R.~A. Iltis, and J.~D. Gibson, ``A qrd-m/kalman filter-based
  detection and channel estimation algorithm for mimo-ofdm systems,''
  \emph{IEEE Transactions on Wireless Communications}, vol.~4, no.~2, pp.
  710--721, March 2005.

\bibitem{li_tc06}
H.~Minn, N.~Al-Dhahir, and Y.~Li, ``Optimal training signals for mimo ofdm
  channel estimation in the presence of frequency offset and phase noise,''
  \emph{IEEE Transactions on Communications}, vol.~54, no.~10, pp. 1754--1759,
  Oct 2006.

\bibitem{gershman_tsp06}
M.~Biguesh and A.~B. Gershman, ``Training-based mimo channel estimation: a
  study of estimator tradeoffs and optimal training signals,'' \emph{IEEE
  Transactions on Signal Processing}, vol.~54, no.~3, pp. 884--893, March 2006.

\bibitem{3gpp_13}
3rd Generation Partnership Project (3GPP~TM), ``{Technical Specification Group
  Radio Access Network; Evolved Universal Terrestrial Radio Access (E-UTRA);
  Physical Channels and Modulation (Release 10) },'' 3GPP TS 36.211 V10.7.0
  (2013-02).

\bibitem{li_socc15}
L.~Li and W.~Zhang, ``{On the Encoding Complexity of Systematic Polar Codes},''
  in \emph{IEEE International System-on-Chip Conference (SOCC)}, Sep 2015, pp.
  508--513.

\bibitem{gross_tc16}
G.~Sarkis, I.~Tal, P.~Giard, A.~Vardy, C.~Thibeault, and W.~J. Gross,
  ``{Flexible and Low-Complexity Encoding and Decoding of Systematic Polar
  Codes},'' \emph{IEEE Transactions on Communication}, vol.~64, no.~7, pp.
  2732--2745, June 2016.

\bibitem{eslami_allerton10}
A.~Eslami and H.~Pishro-Nik, ``{On Bit Error Rate Performance of Polar Codes in
  Finite Regime},'' in \emph{48th Annual Allerton Conference on Communication,
  Control, and Computing (Allerton)}, 2010, pp. 188--194.

\bibitem{eslami_isit11}
------, ``{A Practical Approach to Polar Codes},'' in \emph{IEEE International
  Symposium on Information Theory}, 2011, pp. 16--20.

\bibitem{cho_10}
Y.~S. Cho, J.~Kim, W.~Y. Yang, and C.~G. Kang, \emph{MIMO-OFDM Wireless
  Communications with Matlab}.\hskip 1em plus 0.5em minus 0.4em\relax John
  Wiely \& Songs (Asia) Pte Ltd, 2010.

\end{thebibliography}
\bibliographystyle{IEEEtran}
\begin{IEEEbiography}
[{\includegraphics[width=1in,height=1.25in,clip,keepaspectratio]{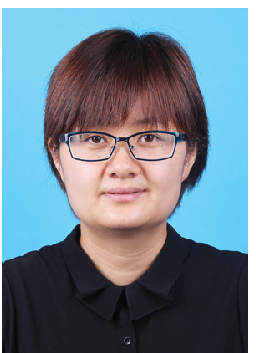}}]{Liping Li}
(S'07-M'15) is now an associate professor of the Key Laboratory of Intelligent Computing and Signal Processing of the Ministry of Education of China, Anhui University, Hefei, China. She got her PhD in Dept. of Electrical and Computer Engineering at North Carolina State University, Raleigh, NC, USA, in 2009. Her current research interest is in channel coding, especially polar codes. Dr. Li's research topic during her PhD studies was multiple-access interference analysis and synchronization for ultra-wideband communications. Then she worked on a LTE indoor channel sounding and modeling project in University of Colorado at Boulder, collaborating with Verizon. From 2010 to 2013, she worked at Maxlinear Inc. as a staff engineer in the communication group. At Maxlinear, she worked on SoC designs for the ISDB-T standard and the DVB-S standard, covering modules on OFDM and LDPC. In Sept. 2013, she joined Anhui University and started her research on polar codes until now.
\end{IEEEbiography}
\begin{IEEEbiography}
[{\includegraphics[width=1in,height=1.25in,clip,keepaspectratio]{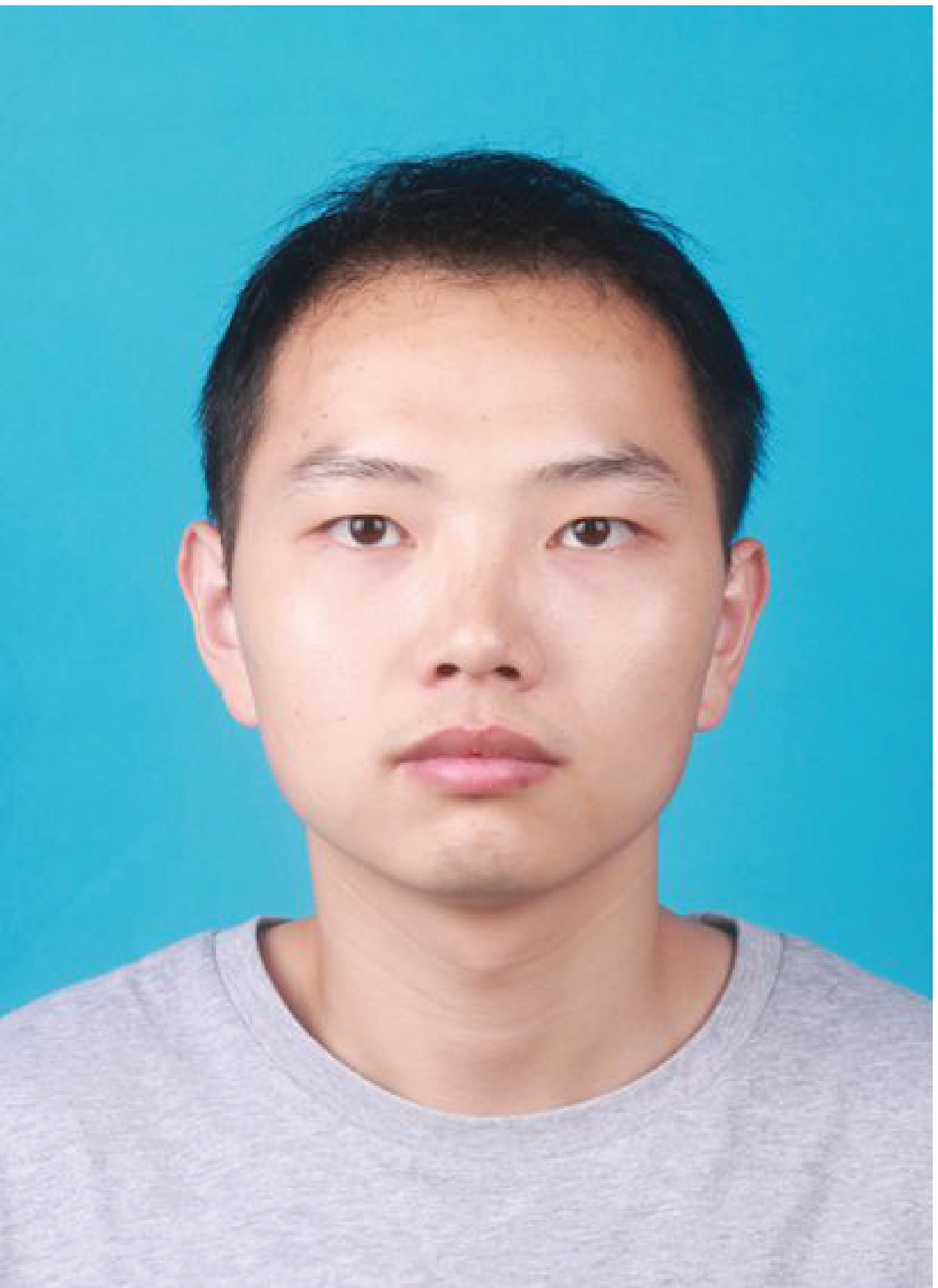}}]
{Zuzheng Xu}
(S'17) is currently pursuing the M.S.
degree in the School of Electronics and Information Engineering, Anhui University. His research interest is in polar codes. Specifically, he is interested in improving  polar code performance in wireless fading channels.
\end{IEEEbiography}
\begin{IEEEbiography}
[{\includegraphics[width=1in,height=1.25in,clip,keepaspectratio]{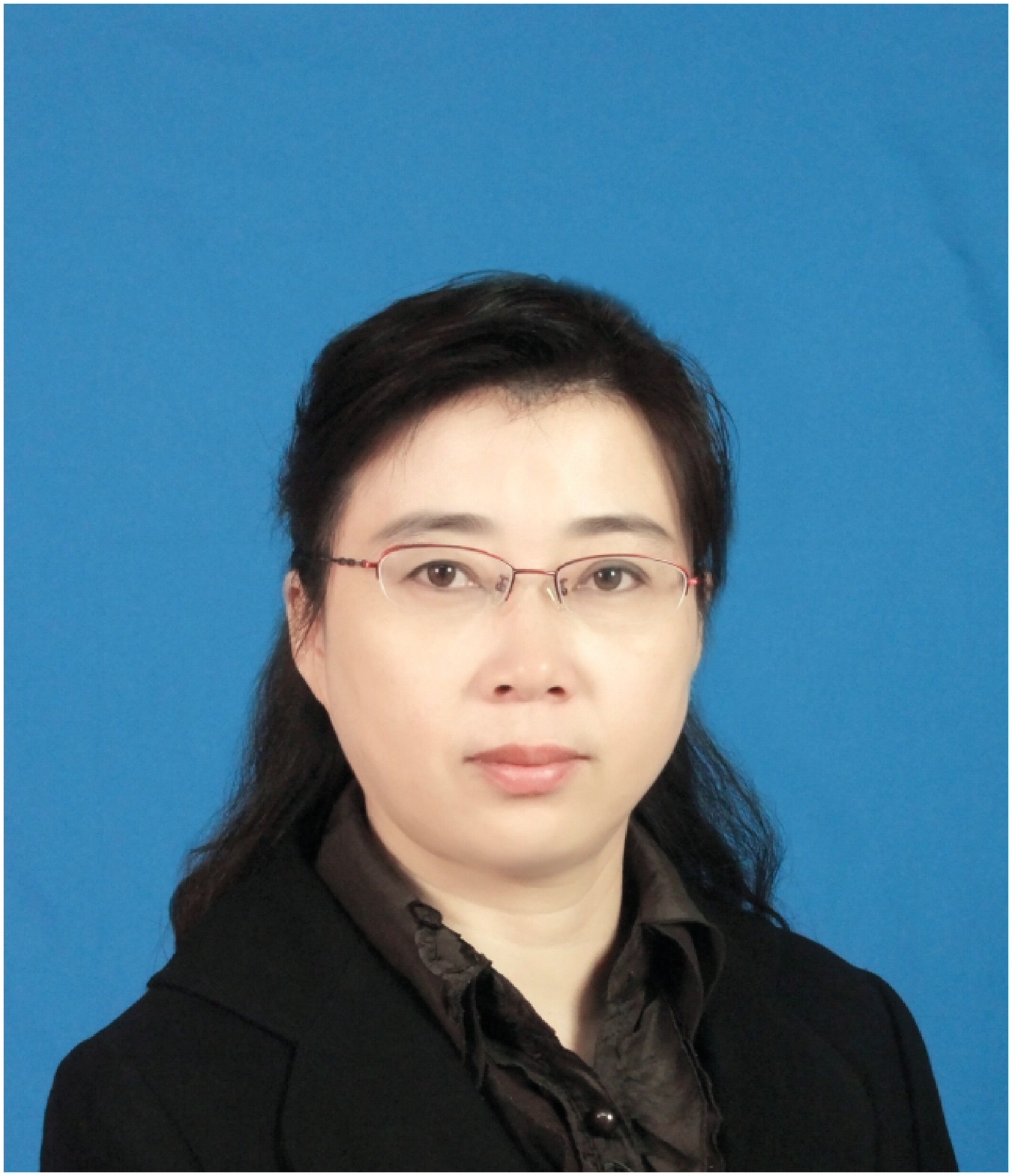}}]
{Yanjun Hu}
(SM'07) is a professor of the Key Laboratory of Intelligent Computing and Signal Processing of the Ministry of Education of China, Anhui University, Hefei, China. She is a senior member of IEEE.
Dr. Hu got her PhD in Communications and Information System in 1998 from University
of Science and Technology of China. . She was a visiting scholar from Feb. 2005 to Aug. 2006 in University of Ottawa, Canada. Dr. Hu's research covers wireless communications and wireless sensor networks. She's published over one hundred research papers and holds three patents. She has funded by NSFC and other key projects in China.
\end{IEEEbiography}
\end{document}